\documentclass[aip,cha,preprint]{revtex4-1}
\usepackage[dvips]{graphicx}
\usepackage{amsmath}
\newcommand{\ud}{\mathrm{d}}
\begin{document}

%\preprint{}

\title{Microwave coupled electron tunneling measurement of Co nanoparticles} %Title of paper



\author{W. Jiang}
\email[]{wjiang31@gatech.edu}
%\thanks{}
\affiliation{School of Physics, Georgia Institute of Technology, 837 State Street, Atlanta, GA 30332}

\author{F. Tijiwa Birk}
%\thanks{}
\affiliation{School of Physics, Georgia Institute of Technology, 837 State Street, Atlanta, GA 30332}

\author{D. Davidovi\'c}
%\email[]{dragomir.davidovic@physics.gatech.edu}
%\thanks{}
\affiliation{School of Physics, Georgia Institute of Technology, 837 State Street, Atlanta, GA 30332}

\date{\today}

\begin{abstract}
We study electron tunneling through Co nanoparticles in the presence of repeated microwave pulses at $4.2$K. While individual pulses are too weak to affect the magnetic switching field, repeated microwave pulses start to reduce the magnetic switching field at $10\mu$s spacing. We use I-V curve as a thermometer to show that the microwave pulses do not heat the  sample, showing that magnetization in Co nanoparticles is directly excited by microwave pulses, and the relaxation time of the excitation energy is in the range of microsecond.
\end{abstract}


\maketitle %\maketitle must follow title, authors, abstract and \pacs

%\section{}
%\label{}
%\subsection{}
%\subsubsection{}
In the past decade, magnetic nanoparticles have generated great interest because of their novel behavior and potential importance in further miniaturization of magnetic storage. Difficult experiments have been carried out to study the electronic and magnetic properties of individual magnetic nanoparticles \cite{lederman,wernsdorfer,orozco,gueron,orozco2,deshmukh,jamet1,thirion,jamet,yakushiji,tamion}. Some of these experiments \cite{gueron,deshmukh} focus on electron tunneling measurements of discrete energy levels of the nanoparticles. Rich phenomena revealed by these experiments, such as hysteretic energy levels shifts in external magnetic field, and abundance of low-energy excitations, motivated a lot of theoretical effort to describe the physics of nanoscale ferromagnetic systems\cite{canali,kleff,waintal,michalak}. These works improved the understanding of ferromagnetism and tunneling transport in nanoparticles. On the other hand, some experiments are performed to directly investigate the magnetization of nanoparticles using magnetic force microscopy\cite{lederman} and microSQUID (micro superconducting quantum interference device)\cite{wernsdorfer,orozco,orozco2,thirion, jamet, tamion}. Radio-frequency (RF) field pulse can reduce the magnetization switching field of a Co nanoparticle, and ferromagnetic resonance at low temperatures can produce a bimodal distribution of the switching field.\cite{thirion, tamion}

In this letter, we present a technique to study the effects of nanosecond microwave pulses on magnetic hysteresis loops of Co nanoparticles,  by electron tunneling.  We confirm that the heating effect on the sample due to the microwave pulses is very weak by using the I-V curve as a thermometer. Thus, the microwave field couples directly to the  magnetization of the nanoparticles. We find that the magnetic energy delivered by individual pulses into the nanoparticle dissipates on the time scale of several microsecond. Our technique combines both electron tunneling and microwave pumping, and can be used to explore magnetic properties from tunnel spectroscopy as well as magnetization dynamics.

\begin{figure}
\includegraphics[width=0.98\textwidth]{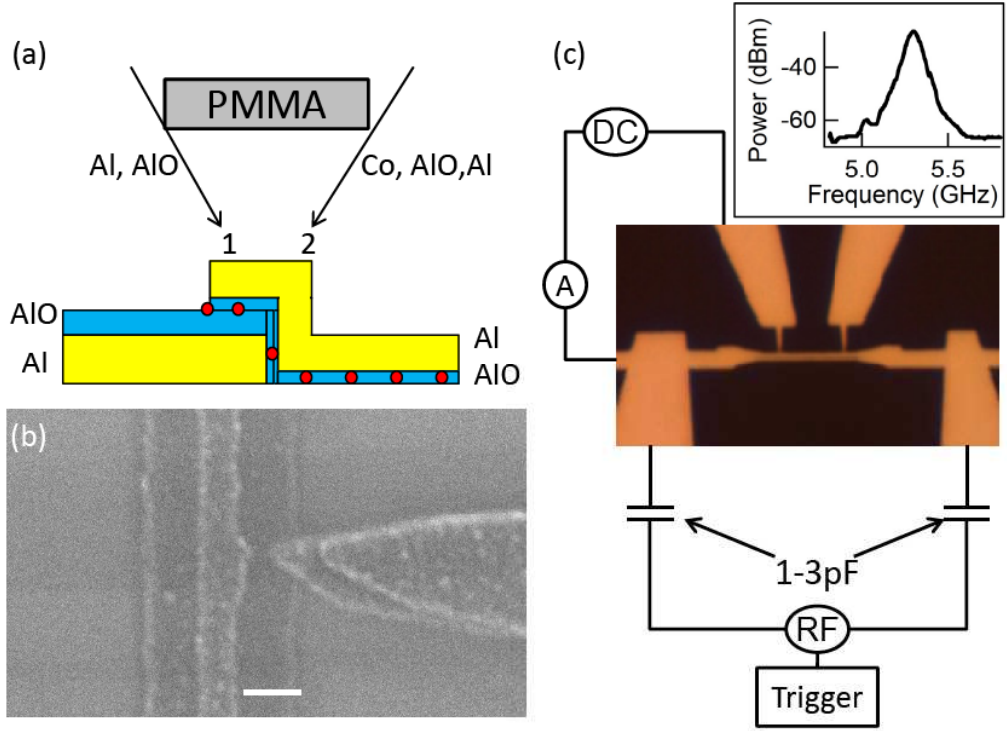}
\caption{(a) Sketch representing sample fabrication process. (b) Scanning electron micrograph of a typical device. The scale bar indicates $0.2\mu$m. (c) Scheme of electrical circuit. Inset: Fourier spectrum of repeated microwave pulses. }
\end{figure}

Our samples are prepared by electron beam lithography (EBL) and shadow evaporation, similar to the technique described previously \cite{davidovic,wei}. First, a Poly(methyl methacrylate) bridge is defined by EBL over SiO$_{2}$ substrate. Next, we deposit 10nm of Al at $2\times10^{-7}$ Torr base pressure along direction 1.  Then,  in the same direction, we deposit $1$-$1.5$nm Al$_{2}$O$_{3}$ by reactive evaporation of Al\cite{davidovic}, at a rate of $0.35$nm/s. The oxygen pressure is kept at $2.5\times10^{-5}$Torr for $10$-$20$s in the evaporation process, so the edge of the Al lead is also slightly oxidized.  After that, we shut down the oxygen flow. When the pressure decreases to $10^{-7}$ Torr range, we rotate the sample by $32^{\circ}$ and deposit $0.5$nm Co along direction 2.  At $0.5$nm thickness, Co forms isolated nanoparticles with $1$-$4$nm in diameter and spaced by $2$-$5$nm\cite{gueron}. Finally, we deposit another $1$-$1.5$nm Al$_{2}$O$_{3}$ layer by reactive evaporation and top it with 10nm Al (Fig. 1(a)).  In each sample, there are two tunnel junctions with Co nanoparticles embedded in Al$_{2}$O$_{3}$ insulating matrix.  The tunnel junction is formed by overlapping the finger like lead and the stripe like lead, as it is shown by Fig. 1(b). In this letter, the number of nanoparticles in each junction is much larger than one. Due to the exponential dependence of the tunnel resistance on barrier thickness, only a few nanoparticles contribute significantly to transport, as will be shown.
The Al strip lead is also used to apply the microwave pulses to Co nanoparticles. Fig. 1(c) displays the circuit scheme of  the experiment.  During the measurements, DC bias voltage is applied to one tunnel junction while the other one is left open. The microwave leads are coupled to the cable via two $1$-$3$pF capacitors to electrically isolate the microwave circuit.

Repeated nanosecond microwave pulses are applied in the experiment. The microwave is generated by Agilent 83640B  at a fixed power output level. The center frequency of the microwave is $5.3$GHz which is in the range used in Ref. \onlinecite{tamion} . The spacing between pulses ($T_{s}$) is varied from $1000\mu$s to $0.33\mu$s using pulse modulation and SR-DS345 arbitrary waveform generator. The average single pulse duration is $\approx 8$ns according to the statistics of real time measurements using microwave power detector and Lecroy 9370 oscilloscope.  The calculation based on the Fourier spectrum of pulses taken by HP 8590E Spectrum Analyzer (inset of Fig. 1(c)) also shows a similar pulse length.

\begin{figure}
\includegraphics[width=0.98\textwidth]{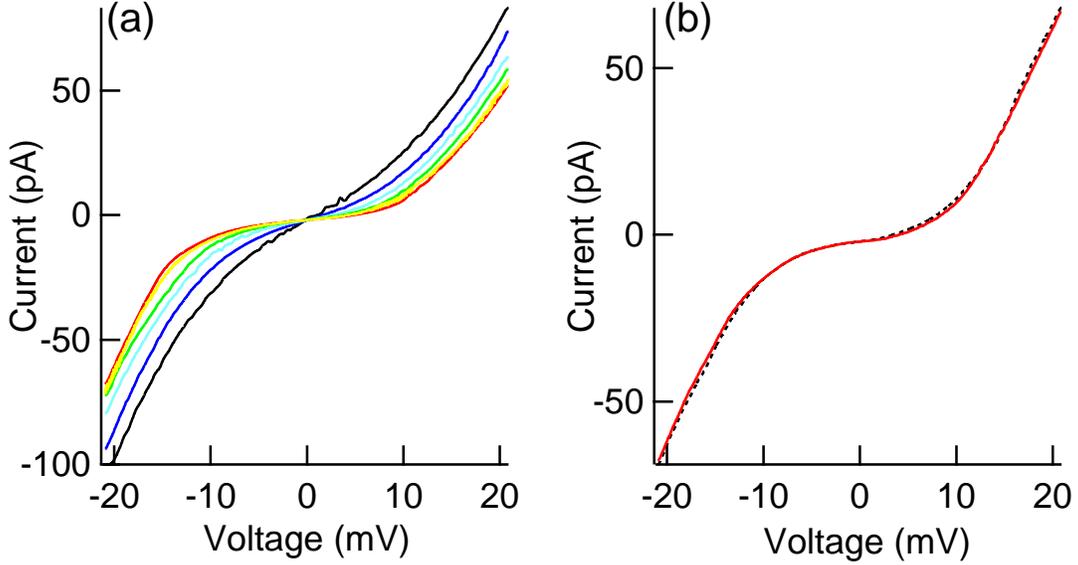}
\caption{(a) Average I-V curves at  $27$K, $16.6$K, $13.1$K, $9.3$K, $5.8$K, and $4.2$K (from top to bottom at positive bias voltage).  (b) Average I-V curves at $T_{s}=$ $1000\mu$s (red/solid line) and $1.25\mu$s (black/dashed line).}
\end{figure}

First, we confirm that the microwave pulses applied have a negligible heating effect on the sample, by measuring the I-V curve of the sample. Fig. 2(a) displays the I-V curves at different temperatures, from $4.2$K to $27$K. At higher temperature, the current at certain bias voltage is  higher than it is at lower temperature. Then,  we measure the I-V curve at $4.2$K under repeated microwave pluses with different $T_{s}$, as shown in Fig. 2(b). Each curve in Fig. 2 is an average of $2$-$5$ voltage cycles \cite{Note1}. The I-V curve at $T_{s}=1.25\mu$s is nearly identical to that at $T_{s}=1000\mu$s, indicating that the heating effect of microwave pulses with spacing equal or larger than $1.25\mu$s  is still negligible.

Next, we measure the tunneling current through nanoparticles versus magnetic field  in presence of repeated nanosecond microwave pulses at $4.2$K .
We fix the bias voltage at $10$mV across the sample while sweeping the magnetic field and the results at different $T_{s}$
are displayed in Fig. 3(a).  We also measure the current hysteresis loops with the same bias voltage at different temperatures for comparison (Fig. 3(b)). Each loop displayed in Fig. 3 is averaged over $8$-$40$ field cycles, while the one at $1000\mu$s spacing is averaged over $700$ cycles. Current loops with finite $T_{s}$ and $7$K are offset down with $4$pA spacing for clarity\cite{Note2}.  The current shifts with magnetic field are hysteretic with respect to the direction of the field sweep. The magnetoresistance (MR) in our samples is attributed to the magnetic field dependence of the discrete energy levels. This dependence has been studied in Refs. \onlinecite{gueron, deshmukh} where the magnetic anisotropy energy was found to have a significant contribution.  We have confirmed such dependence of the levels in our sample at dilution refrigerator temperatures\cite{birk}.  The dependence of the levels on the applied magnetic field leads to MR at low temperature where the levels are well resolved. At $4.2$K, energy levels broaden but some MR still remains\cite{birk}. In addition to the field dependence of the levels due to the magnetic anisotropy, magneto-Coulomb effect\cite{ono, shimada, ono2, molen, mantel} may also contribute to the MR in our devices. So, the transition from low(high) current to high(low) current state during the sweep of magnetic field suggests magnetization reversal in Co nanoparticles. The width of the magnetic transitions in Fig. 3 arises from averaging of the hysteresis loops over many field cycles. In individual hysteresis loops, the magnetic transitions are discontinuous, and switching field of the transitions fluctuates among different field cycles\cite{Note2}. Since different nanoparticles are likely to have transitions at different fields, the small number of magnetic transitions shows  that the number of nanoparticles involved in transport is small.  One transition indicated by cross in Fig. 3(a) is observed in all hysteresis loops. Another transition indicated by star in Fig. 3(a) is also clear except in the loop at $T_{s}=1.25\mu$s. 

\begin{figure}
\includegraphics[width=0.98\textwidth]{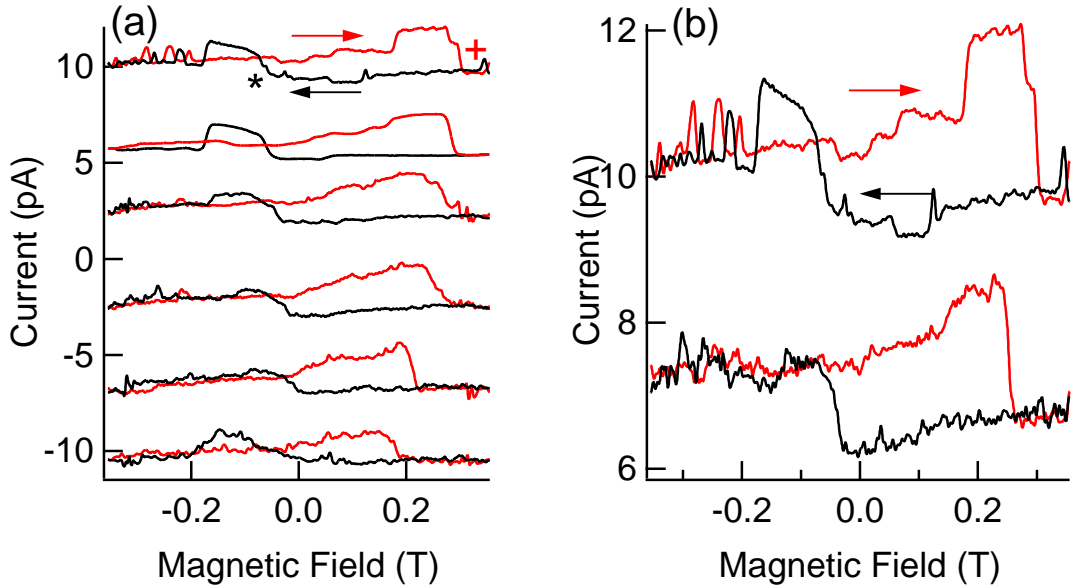}
\caption{(a) Average current loops at $4.2$K when $T_{s}=$ $\infty$ (no microwave), $1000\mu$s, $10\mu$s, $5\mu$s, $2.5\mu$s, and $1.25\mu$s (top to bottom). The right and left arrow corresponds to increasing and decreasing magnetic field respectively. (b) Average current loops at $4.2$K and $7$K (top to bottom). Current loops with finite $T_{s}$ in (a) and $7$K in (b) are offset down with $4$pA spacing.}
\end{figure}

When $T_{s}=1000\mu$s, the hysteresis loop is nearly identical to that without microwaves, showing that a single microwave pulse is too weak to  magnetically excite the Co nanoparticles. Reductions in switching field of the magnetic transitions are observed when the spacing is below $10\mu$s, suggesting microwave triggered switch \cite{thirion,tamion} (Fig. 3(a)). Fig. 3(b) shows the similar reduction effect due to thermal fluctuation\cite {neel, brown}. At $1.25\mu$s spacing, the switching field of the transition indicated by cross is reduced by $\approx40$\% compared  with that at $1000\mu$s.  The reduction of the switching field caused by a temperature increase from $4.2$K to $7$K is much weaker than $40$\%. The electron temperature in the leads at $T_{s}=1.25\mu$s is estimated to be $\le 4.4$K from the I-V curve (Fig. 2(b)) .  Similarly, the decrease in the switching field of the transition indicated by star when $T_{s}=2.5\mu$s  is larger than that at $7$K. %At $T_{s}=1.25\mu$ and $2.5\mu$s, the electron temprature  is much lower than $7$K,  because we obtain $\approx 6.3$K for the electron temperature at $T_{s}=0.33\mu$s which leads to a much higher microwave power.
These results demonstrate that magnetic excitations in Co nanoparticles are pumped directly by microwave pulses, not by ordinary sample heating. The reduction of  switching field  is still observable at $10\mu$s pulse spacing, which implies that the magnetic energy delivered by a single pulse does not  completely dissipate in $10\mu$s. This suggests that the spin relaxation time in our Co nanoparticles is in the order of microsecond. Long lower bounds of the spin-relaxation time $T_1$ in Co nanoparticles have been observed before.\cite{deshmukh, yakushiji,birk} The work in this letter narrows down $T_1$ at 4.2K to the microsecond range.

We can use the I-V curve to estimate the magnetic field at the nanoparticle due to the RF current. The broadening of the I-V curve by the microwave pulses becomes visible in our measurement when the spacing between pulses is $0.33\mu$s (not shown). We also induce broadening of the I-V curve by injection of a continuous AC current at frequency $1.2$kHz, instead of microwave pulses (after shorting the capacitors). The amplitude of the AC current $I_{ac}$ can easily be determined using a standard lock-in technique. We use a low frequency transformer to isolate the AC circuit. Then we measure the I-V curve at different AC current. By matching the I-V curve at different $T_{s}$ and $I_{ac}$, we can find the power of microwave pulses with certain $T_{s}$ from the amplitude of corresponding $I_{ac}$. Based on this result, we can solve the equations, $I(t) =\int_{-\infty}^{\infty} \, I(f)e^{\imath 2\pi ft} \, \ud f$, $P(t) =RI^{2}(t)$, and $\int^{\infty}_{-\infty} \, P(t) \, \ud t/T_{s}  =RI^{2}_{ac}$,
%\begin{flalign}
%\qquad I(t)& =\int_{-\infty}^{\infty} \, I(f)e^{\imath 2\pi ft} \, \ud f &
%\end{flalign}
%\begin{flalign}
%\qquad P(t)& =RI^{2}(t)&
%\end{flalign}
%\begin{flalign}
%\qquad  \frac{\int^{\infty}_{-\infty} \, P(t) \, \ud t}{T_{s}} & =RI^{2}_{ac} &
%\end{flalign}
where $I(f)$ is the current density of frequency and $P(t)$ is the power of a single pulse. Under the assumption that  $|I(f)|^2$ is Gaussian distributed according to the Fourier spectrum, we obtain  the expression for microwave induced current $I(t)$ in time domain after Fourier transform of $I(f)$. Then, we can find the amplitude of $I(t)$ by plugging this
expression into  the second equation. The amplitude of the RF-field on the nanoparticle is estimated to be $\approx 5\times 10^{-4}$T from the lead geometry and $I(t)$. If a white noise instead of microwave pulses is coupled to the sample, the damping parameter  of Co nanoparticles could be estimated by corresponding  the applied random magnetic field  with the one obtained from magnetic thermal noise\cite{brown}. This will be the subject of future work.

In summary, we measure the hysteresis loops of Co nanoparticles in presence of nanosecond microwave pulses by electron tunneling and demonstrate that microwave pulses can  pump magnetic excitations in these nanoparticles directly, without heating the leads. The magnetic switching field is reduced by repeated pulses with spacing $\le 10\mu$s, suggesting that magnetization relaxation time is in the range of microsecond.   The method we used in the experiment provides a possible way to measure the magnetic damping parameter  and magnetization dynamics of Co nanoparticles.

% If in two-column mode, this environment will change to single-column format so that long equations can be displayed.
%\begin{widetext}
%$$\mbox{put long equation here}$$
%\end{widetext}

% Figures should be put into the text as floats.
% Use the graphics or graphicx packages (distributed with LaTeX2e).
% Put the label that you will use with \ref{} command in the braces of the \label{} command.
%
% \begin{figure}
% \end{figure}

% \begin{table}

% \end{tabular}
% \end{table}

\begin{acknowledgments}
This work has been supported by the Department of Energy (DE-FG02-06ER46281). We thank M. S. Chapman and C. D. Hamley for the help  with the microwave pulses characterization. We thank C. E. Malec for valuable discussions.

\end{acknowledgments}

%\bibliography{apl}

\begin{thebibliography}{0}%
\makeatletter
\providecommand \@ifxundefined [1]{%
 \@ifx{#1\undefined}
}%
\providecommand \@ifnum [1]{%
 \ifnum #1\expandafter \@firstoftwo
 \else \expandafter \@secondoftwo
 \fi
}%
\providecommand \@ifx [1]{%
 \ifx #1\expandafter \@firstoftwo
 \else \expandafter \@secondoftwo
 \fi
}%
\providecommand \natexlab [1]{#1}%
\providecommand \enquote  [1]{``#1''}%
\providecommand \bibnamefont  [1]{#1}%
\providecommand \bibfnamefont [1]{#1}%
\providecommand \citenamefont [1]{#1}%
\providecommand \href@noop [0]{\@secondoftwo}%
\providecommand \href [0]{\begingroup \@sanitize@url \@href}%
\providecommand \@href[1]{\@@startlink{#1}\@@href}%
\providecommand \@@href[1]{\endgroup#1\@@endlink}%
\providecommand \@sanitize@url [0]{\catcode `\\12\catcode `\$12\catcode
  `\&12\catcode `\#12\catcode `\^12\catcode `\_12\catcode `\%12\relax}%
\providecommand \@@startlink[1]{}%
\providecommand \@@endlink[0]{}%
\providecommand \url  [0]{\begingroup\@sanitize@url \@url }%
\providecommand \@url [1]{\endgroup\@href {#1}{\urlprefix }}%
\providecommand \urlprefix  [0]{URL }%
\providecommand \Eprint [0]{\href }%
\providecommand \doibase [0]{http://dx.doi.org/}%
\providecommand \selectlanguage [0]{\@gobble}%
\providecommand \bibinfo  [0]{\@secondoftwo}%
\providecommand \bibfield  [0]{\@secondoftwo}%
\providecommand \translation [1]{[#1]}%
\providecommand \BibitemOpen [0]{}%
\providecommand \bibitemStop [0]{}%
\providecommand \bibitemNoStop [0]{.\EOS\space}%
\providecommand \EOS [0]{\spacefactor3000\relax}%
\providecommand \BibitemShut  [1]{\csname bibitem#1\endcsname}%
\let\auto@bib@innerbib\@empty
%</preamble>
\end{thebibliography}%


\begin{thebibliography}{27}%
\makeatletter
\providecommand \@ifxundefined [1]{%
 \@ifx{#1\undefined}
}%
\providecommand \@ifnum [1]{%
 \ifnum #1\expandafter \@firstoftwo
 \else \expandafter \@secondoftwo
 \fi
}%
\providecommand \@ifx [1]{%
 \ifx #1\expandafter \@firstoftwo
 \else \expandafter \@secondoftwo
 \fi
}%
\providecommand \natexlab [1]{#1}%
\providecommand \enquote  [1]{``#1''}%
\providecommand \bibnamefont  [1]{#1}%
\providecommand \bibfnamefont [1]{#1}%
\providecommand \citenamefont [1]{#1}%
\providecommand \href@noop [0]{\@secondoftwo}%
\providecommand \href [0]{\begingroup \@sanitize@url \@href}%
\providecommand \@href[1]{\@@startlink{#1}\@@href}%
\providecommand \@@href[1]{\endgroup#1\@@endlink}%
\providecommand \@sanitize@url [0]{\catcode `\\12\catcode `\$12\catcode
  `\&12\catcode `\#12\catcode `\^12\catcode `\_12\catcode `\%12\relax}%
\providecommand \@@startlink[1]{}%
\providecommand \@@endlink[0]{}%
\providecommand \url  [0]{\begingroup\@sanitize@url \@url }%
\providecommand \@url [1]{\endgroup\@href {#1}{\urlprefix }}%
\providecommand \urlprefix  [0]{URL }%
\providecommand \Eprint [0]{\href }%
\providecommand \doibase [0]{http://dx.doi.org/}%
\providecommand \selectlanguage [0]{\@gobble}%
\providecommand \bibinfo  [0]{\@secondoftwo}%
\providecommand \bibfield  [0]{\@secondoftwo}%
\providecommand \translation [1]{[#1]}%
\providecommand \BibitemOpen [0]{}%
\providecommand \bibitemStop [0]{}%
\providecommand \bibitemNoStop [0]{.\EOS\space}%
\providecommand \EOS [0]{\spacefactor3000\relax}%
\providecommand \BibitemShut  [1]{\csname bibitem#1\endcsname}%
\let\auto@bib@innerbib\@empty
%</preamble>
\bibitem [{\citenamefont {Lederman}, \citenamefont {Schultz},\ and\
  \citenamefont {Ozaki}(1994)}]{lederman}%
  \BibitemOpen
  \bibfield  {author} {\bibinfo {author} {\bibfnamefont {M.}~\bibnamefont
  {Lederman}}, \bibinfo {author} {\bibfnamefont {S.}~\bibnamefont {Schultz}}, \
  and\ \bibinfo {author} {\bibfnamefont {M.}~\bibnamefont {Ozaki}},\ }\href
  {\doibase 10.1103/PhysRevLett.73.1986} {\bibfield  {journal} {\bibinfo
  {journal} {Phys. Rev. Lett.}\ }\textbf {\bibinfo {volume} {73}},\ \bibinfo
  {pages} {1986} (\bibinfo {year} {1994})}\BibitemShut {NoStop}%
\bibitem [{\citenamefont {Wernsdorfer}\ \emph {et~al.}(1997)\citenamefont
  {Wernsdorfer}, \citenamefont {Orozco}, \citenamefont {Hasselbach},
  \citenamefont {Benoit}, \citenamefont {Barbara}, \citenamefont {Demoncy},
  \citenamefont {Loiseau}, \citenamefont {Pascard},\ and\ \citenamefont
  {Mailly}}]{wernsdorfer}%
  \BibitemOpen
  \bibfield  {author} {\bibinfo {author} {\bibfnamefont {W.}~\bibnamefont
  {Wernsdorfer}}, \bibinfo {author} {\bibfnamefont {E.~B.}\ \bibnamefont
  {Orozco}}, \bibinfo {author} {\bibfnamefont {K.}~\bibnamefont {Hasselbach}},
  \bibinfo {author} {\bibfnamefont {A.}~\bibnamefont {Benoit}}, \bibinfo
  {author} {\bibfnamefont {B.}~\bibnamefont {Barbara}}, \bibinfo {author}
  {\bibfnamefont {N.}~\bibnamefont {Demoncy}}, \bibinfo {author} {\bibfnamefont
  {A.}~\bibnamefont {Loiseau}}, \bibinfo {author} {\bibfnamefont
  {H.}~\bibnamefont {Pascard}}, \ and\ \bibinfo {author} {\bibfnamefont
  {D.}~\bibnamefont {Mailly}},\ }\href {\doibase 10.1103/PhysRevLett.78.1791}
  {\bibfield  {journal} {\bibinfo  {journal} {Phys. Rev. Lett.}\ }\textbf
  {\bibinfo {volume} {78}},\ \bibinfo {pages} {1791} (\bibinfo {year}
  {1997})}\BibitemShut {NoStop}%
\bibitem [{\citenamefont {Orozco}\ \emph {et~al.}(1998)\citenamefont {Orozco},
  \citenamefont {Wernsdorfer}, \citenamefont {Barbara}, \citenamefont
  {Hasselbach}, \citenamefont {Benoit},\ and\ \citenamefont {Mailly}}]{orozco}%
  \BibitemOpen
  \bibfield  {author} {\bibinfo {author} {\bibfnamefont {E.}~\bibnamefont
  {Orozco}}, \bibinfo {author} {\bibfnamefont {W.}~\bibnamefont {Wernsdorfer}},
  \bibinfo {author} {\bibfnamefont {B.}~\bibnamefont {Barbara}}, \bibinfo
  {author} {\bibfnamefont {K.}~\bibnamefont {Hasselbach}}, \bibinfo {author}
  {\bibfnamefont {A.}~\bibnamefont {Benoit}}, \ and\ \bibinfo {author}
  {\bibfnamefont {D.}~\bibnamefont {Mailly}},\ }\href@noop {} {\bibfield
  {journal} {\bibinfo  {journal} {Magnetics, IEEE Transactions on}\ }\textbf
  {\bibinfo {volume} {34}},\ \bibinfo {pages} {979} (\bibinfo {year}
  {1998})}\BibitemShut {NoStop}%
\bibitem [{\citenamefont {Gu\'eron}\ \emph {et~al.}(1999)\citenamefont
  {Gu\'eron}, \citenamefont {Deshmukh}, \citenamefont {Myers},\ and\
  \citenamefont {Ralph}}]{gueron}%
  \BibitemOpen
  \bibfield  {author} {\bibinfo {author} {\bibfnamefont {S.}~\bibnamefont
  {Gu\'eron}}, \bibinfo {author} {\bibfnamefont {M.~M.}\ \bibnamefont
  {Deshmukh}}, \bibinfo {author} {\bibfnamefont {E.~B.}\ \bibnamefont {Myers}},
  \ and\ \bibinfo {author} {\bibfnamefont {D.~C.}\ \bibnamefont {Ralph}},\
  }\href@noop {} {\bibfield  {journal} {\bibinfo  {journal} {Phys. Rev. Lett.}\
  }\textbf {\bibinfo {volume} {83}},\ \bibinfo {pages} {4148} (\bibinfo {year}
  {1999})}\BibitemShut {NoStop}%
\bibitem [{\citenamefont {Orozco}\ \emph {et~al.}(2000)\citenamefont {Orozco},
  \citenamefont {Wernsdorfer}, \citenamefont {Barbara}, \citenamefont
  {Beno{\^{i}}t}, \citenamefont {Mailly},\ and\ \citenamefont
  {Thiaville}}]{orozco2}%
  \BibitemOpen
  \bibfield  {author} {\bibinfo {author} {\bibfnamefont {E.~B.}\ \bibnamefont
  {Orozco}}, \bibinfo {author} {\bibfnamefont {W.}~\bibnamefont {Wernsdorfer}},
  \bibinfo {author} {\bibfnamefont {B.}~\bibnamefont {Barbara}}, \bibinfo
  {author} {\bibfnamefont {A.}~\bibnamefont {Benoit}}, \bibinfo {author}
  {\bibfnamefont {D.}~\bibnamefont {Mailly}}, \ and\ \bibinfo {author}
  {\bibfnamefont {A.}~\bibnamefont {Thiaville}},\ }\href@noop {} {\bibfield
  {journal} {\bibinfo  {journal} {J. Appl. Phys}\ }\textbf
  {\bibinfo {volume} {87}},\ \bibinfo {pages} {5097} (\bibinfo {year}
  {2000})}\BibitemShut {NoStop}%
\bibitem [{\citenamefont {Deshmukh}\ \emph {et~al.}(2001)\citenamefont
  {Deshmukh}, \citenamefont {Kleff}, \citenamefont {Gu{\'{e}}ron},
  \citenamefont {Orozco}, \citenamefont {Pasupathy}, \citenamefont {von
  Delft},\ and\ \citenamefont {Ralph}}]{deshmukh}%
  \BibitemOpen
  \bibfield  {author} {\bibinfo {author} {\bibfnamefont {M.~M.}\ \bibnamefont
  {Deshmukh}}, \bibinfo {author} {\bibfnamefont {S.}~\bibnamefont {Kleff}},
  \bibinfo {author} {\bibfnamefont {S.}~\bibnamefont {Gu{\'{e}}ron}}, \bibinfo
  {author} {\bibfnamefont {E.~B.}\ \bibnamefont {Orozco}}, \bibinfo {author}
  {\bibfnamefont {A.~N.}\ \bibnamefont {Pasupathy}}, \bibinfo {author}
  {\bibfnamefont {J.}~\bibnamefont {von Delft}}, \ and\ \bibinfo {author}
  {\bibfnamefont {D.~C.}\ \bibnamefont {Ralph}},\ }\href@noop {} {\bibfield
  {journal} {\bibinfo  {journal} {Phys. Rev. Lett.}\ }\textbf {\bibinfo
  {volume} {87}},\ \bibinfo {pages} {226801} (\bibinfo {year}
  {2001})}\BibitemShut {NoStop}%
\bibitem [{\citenamefont {Jamet}\ \emph {et~al.}(2001)\citenamefont {Jamet},
  \citenamefont {Wernsdorfer}, \citenamefont {Thirion}, \citenamefont {Mailly},
  \citenamefont {Dupuis}, \citenamefont {M\'elinon},\ and\ \citenamefont
  {P\'erez}}]{jamet1}%
  \BibitemOpen
  \bibfield  {author} {\bibinfo {author} {\bibfnamefont {M.}~\bibnamefont
  {Jamet}}, \bibinfo {author} {\bibfnamefont {W.}~\bibnamefont {Wernsdorfer}},
  \bibinfo {author} {\bibfnamefont {C.}~\bibnamefont {Thirion}}, \bibinfo
  {author} {\bibfnamefont {D.}~\bibnamefont {Mailly}}, \bibinfo {author}
  {\bibfnamefont {V.}~\bibnamefont {Dupuis}}, \bibinfo {author} {\bibfnamefont
  {P.}~\bibnamefont {M\'elinon}}, \ and\ \bibinfo {author} {\bibfnamefont
  {A.}~\bibnamefont {P\'erez}},\ }\href@noop {} {\bibfield  {journal} {\bibinfo
   {journal} {J. Magn. Magn. Mater.}\ }\textbf {\bibinfo {volume} {226}},\
  \bibinfo {pages} {1833} (\bibinfo {year} {2001})}\BibitemShut {NoStop}%
\bibitem [{\citenamefont {Thirion}, \citenamefont {Wernsdorfer},\ and\
  \citenamefont {Mailly}(2003)}]{thirion}%
  \BibitemOpen
  \bibfield  {author} {\bibinfo {author} {\bibfnamefont {C.}~\bibnamefont
  {Thirion}}, \bibinfo {author} {\bibfnamefont {W.}~\bibnamefont
  {Wernsdorfer}}, \ and\ \bibinfo {author} {\bibfnamefont {D.}~\bibnamefont
  {Mailly}},\ }\href@noop {} {\bibfield  {journal} {\bibinfo  {journal} {Nature
  Mater.}\ }\textbf {\bibinfo {volume} {2}},\ \bibinfo {pages} {524} (\bibinfo
  {year} {2003})}\BibitemShut {NoStop}%
\bibitem [{\citenamefont {Jamet}\ \emph {et~al.}(2004)\citenamefont {Jamet},
  \citenamefont {Wernsdorfer}, \citenamefont {Thirion}, \citenamefont {Dupuis},
  \citenamefont {M\'elinon},\ and\ \citenamefont {P\'erez}}]{jamet}%
  \BibitemOpen
  \bibfield  {author} {\bibinfo {author} {\bibfnamefont {M.}~\bibnamefont
  {Jamet}}, \bibinfo {author} {\bibfnamefont {W.}~\bibnamefont {Wernsdorfer}},
  \bibinfo {author} {\bibfnamefont {C.}~\bibnamefont {Thirion}}, \bibinfo
  {author} {\bibfnamefont {V.}~\bibnamefont {Dupuis}}, \bibinfo {author}
  {\bibfnamefont {P.}~\bibnamefont {M\'elinon}}, \ and\ \bibinfo {author}
  {\bibfnamefont {A.}~\bibnamefont {P\'erez}},\ }\href@noop {} {\bibfield
  {journal} {\bibinfo  {journal} {Phys. Rev. B}\ }\textbf {\bibinfo {volume}
  {69}},\ \bibinfo {pages} {024401} (\bibinfo {year} {2004})}\BibitemShut
  {NoStop}%
\bibitem [{\citenamefont {Yakushiji}\ \emph {et~al.}(2005)\citenamefont
  {Yakushiji}, \citenamefont {Ernult}, \citenamefont {Imamura}, \citenamefont
  {Yamane}, \citenamefont {Mitani}, \citenamefont {Takanashi}, \citenamefont
  {Maekawa},\ and\ \citenamefont {Fujimori}}]{yakushiji}%
  \BibitemOpen
  \bibfield  {author} {\bibinfo {author} {\bibfnamefont {K.}~\bibnamefont
  {Yakushiji}}, \bibinfo {author} {\bibfnamefont {F.}~\bibnamefont {Ernult}},
  \bibinfo {author} {\bibfnamefont {H.}~\bibnamefont {Imamura}}, \bibinfo
  {author} {\bibfnamefont {K.}~\bibnamefont {Yamane}}, \bibinfo {author}
  {\bibfnamefont {S.}~\bibnamefont {Mitani}}, \bibinfo {author} {\bibfnamefont
  {K.}~\bibnamefont {Takanashi}}, \bibinfo {author} {\bibfnamefont
  {S.}~\bibnamefont {Maekawa}}, \ and\ \bibinfo {author} {\bibfnamefont
  {H.}~\bibnamefont {Fujimori}},\ }\href@noop {} {\bibfield  {journal}
  {\bibinfo  {journal} {Nature Mater.}\ }\textbf {\bibinfo {volume} {4}},\
  \bibinfo {pages} {57} (\bibinfo {year} {2005})}\BibitemShut {NoStop}%
\bibitem [{\citenamefont {Tamion}\ \emph {et~al.}(2010)\citenamefont {Tamion},
  \citenamefont {Raufast}, \citenamefont {Orozco}, \citenamefont {Dupuis},
  \citenamefont {Fournier}, \citenamefont {Crozes}, \citenamefont {Bernstein},\
  and\ \citenamefont {Wernsdorfer}}]{tamion}%
  \BibitemOpen
  \bibfield  {author} {\bibinfo {author} {\bibfnamefont {A.}~\bibnamefont
  {Tamion}}, \bibinfo {author} {\bibfnamefont {C.}~\bibnamefont {Raufast}},
  \bibinfo {author} {\bibfnamefont {E.~B.}\ \bibnamefont {Orozco}}, \bibinfo
  {author} {\bibfnamefont {V.}~\bibnamefont {Dupuis}}, \bibinfo {author}
  {\bibfnamefont {T.}~\bibnamefont {Fournier}}, \bibinfo {author}
  {\bibfnamefont {T.}~\bibnamefont {Crozes}}, \bibinfo {author} {\bibfnamefont
  {E.}~\bibnamefont {Bernstein}}, \ and\ \bibinfo {author} {\bibfnamefont
  {W.}~\bibnamefont {Wernsdorfer}},\ }\href@noop {} {\bibfield  {journal}
  {\bibinfo  {journal} {J. Magn. Magn. Mater.}\ }\textbf {\bibinfo {volume}
  {322}},\ \bibinfo {pages} {1315} (\bibinfo {year} {2010})}\BibitemShut
  {NoStop}%
\bibitem [{\citenamefont {Canali}\ and\ \citenamefont
  {MacDonald}(2000)}]{canali}%
  \BibitemOpen
  \bibfield  {author} {\bibinfo {author} {\bibfnamefont {C.~M.}\ \bibnamefont
  {Canali}}\ and\ \bibinfo {author} {\bibfnamefont {A.~H.}\ \bibnamefont
  {MacDonald}},\ }\href@noop {} {\bibfield  {journal} {\bibinfo  {journal}
  {Phys. Rev. Lett.}\ }\textbf {\bibinfo {volume} {85}},\ \bibinfo {pages}
  {5623} (\bibinfo {year} {2000})}\BibitemShut {NoStop}%
\bibitem [{\citenamefont {Kleff}\ \emph {et~al.}(2001)\citenamefont {Kleff},
  \citenamefont {von Delft}, \citenamefont {Deshmukh},\ and\ \citenamefont
  {Ralph}}]{kleff}%
  \BibitemOpen
  \bibfield  {author} {\bibinfo {author} {\bibfnamefont {S.}~\bibnamefont
  {Kleff}}, \bibinfo {author} {\bibfnamefont {J.}~\bibnamefont {von Delft}},
  \bibinfo {author} {\bibfnamefont {M.~M.}\ \bibnamefont {Deshmukh}}, \ and\
  \bibinfo {author} {\bibfnamefont {D.~C.}\ \bibnamefont {Ralph}},\ }\href@noop
  {} {\bibfield  {journal} {\bibinfo  {journal} {Phys. Rev. B}\ }\textbf
  {\bibinfo {volume} {64}},\ \bibinfo {pages} {220401} (\bibinfo {year}
  {2001})}\BibitemShut {NoStop}%
\bibitem [{\citenamefont {Waintal}\ and\ \citenamefont
  {Parcollet}(2005)}]{waintal}%
  \BibitemOpen
  \bibfield  {author} {\bibinfo {author} {\bibfnamefont {X.}~\bibnamefont
  {Waintal}}\ and\ \bibinfo {author} {\bibfnamefont {O.}~\bibnamefont
  {Parcollet}},\ }\href@noop {} {\bibfield  {journal} {\bibinfo  {journal}
  {Phys. Rev. Lett.}\ }\textbf {\bibinfo {volume} {94}},\ \bibinfo {pages}
  {247206} (\bibinfo {year} {2005})}\BibitemShut {NoStop}%
\bibitem [{\citenamefont {Michalak}\ and\ \citenamefont
  {Canali}(2006)}]{michalak}%
  \BibitemOpen
  \bibfield  {author} {\bibinfo {author} {\bibfnamefont {L.}~\bibnamefont
  {Michalak}}\ and\ \bibinfo {author} {\bibfnamefont {C.~M.}\ \bibnamefont
  {Canali}},\ }\href@noop {} {\bibfield  {journal} {\bibinfo  {journal} {Phys.
  Rev. Lett.}\ }\textbf {\bibinfo {volume} {97}},\ \bibinfo {pages} {096804}
  (\bibinfo {year} {2006})}\BibitemShut {NoStop}%
\bibitem [{\citenamefont {Davidovi\'c}\ and\ \citenamefont
  {Tinkham}(1999)}]{davidovic}%
  \BibitemOpen
  \bibfield  {author} {\bibinfo {author} {\bibfnamefont {D.}~\bibnamefont
  {Davidovi\'c}}\ and\ \bibinfo {author} {\bibfnamefont {M.}~\bibnamefont
  {Tinkham}},\ }\href@noop {} {\bibfield  {journal} {\bibinfo  {journal} {Phys.
  Rev. Lett.}\ }\textbf {\bibinfo {volume} {83}},\ \bibinfo {pages} {1644}
  (\bibinfo {year} {1999})}\BibitemShut {NoStop}%
\bibitem [{\citenamefont {Wei}, \citenamefont {Malec},\ and\ \citenamefont
  {Davidovi\'c}(2007)}]{wei}%
  \BibitemOpen
  \bibfield  {author} {\bibinfo {author} {\bibfnamefont {Y.~G.}\ \bibnamefont
  {Wei}}, \bibinfo {author} {\bibfnamefont {C.~E.}\ \bibnamefont {Malec}}, \
  and\ \bibinfo {author} {\bibfnamefont {D.}~\bibnamefont {Davidovi\'c}},\
  }\href@noop {} {\bibfield  {journal} {\bibinfo  {journal} {Phys. Rev. B}\
  }\textbf {\bibinfo {volume} {76}},\ \bibinfo {pages} {195327} (\bibinfo
  {year} {2007})}\BibitemShut {NoStop}%
\bibitem [{Note1()}]{Note1}%
  \BibitemOpen
  \bibinfo {note} {See supplementary material at for
  \protect \emph {Raw data of Fig. 2}.}\BibitemShut {Stop}%
\bibitem [{Note2()}]{Note2}%
  \BibitemOpen
  \bibinfo {note} {See supplementary material at for
  \protect \emph {Raw data of Fig. 3}.}\BibitemShut {Stop}%
\bibitem [{\citenamefont {Birk}, \citenamefont {Jiang},\ and\ \citenamefont
  {Davidovi\'c}()}]{birk}%
  \BibitemOpen
  \bibfield  {author} {\bibinfo {author} {\bibfnamefont {F.}~\bibnamefont
  {Birk}}, \bibinfo {author} {\bibfnamefont {W.}~\bibnamefont {Jiang}}, \ and\
  \bibinfo {author} {\bibfnamefont {D.}~\bibnamefont {Davidovi\'c}},\
  }\href@noop {} {\enquote {\bibinfo {title} {Pending},}\ }\bibinfo {note}
  {{Pending} (unpublished)}\BibitemShut {NoStop}%
\bibitem [{\citenamefont {Ono}, \citenamefont {Shimada},\ and\ \citenamefont
  {Ootuka}(1997)}]{ono}%
  \BibitemOpen
  \bibfield  {author} {\bibinfo {author} {\bibfnamefont {K.}~\bibnamefont
  {Ono}}, \bibinfo {author} {\bibfnamefont {H.}~\bibnamefont {Shimada}}, \ and\
  \bibinfo {author} {\bibfnamefont {Y.}~\bibnamefont {Ootuka}},\ }\href@noop {}
  {\bibfield  {journal} {\bibinfo  {journal} {J. Phys. Soc. Jpn}\ }\textbf
  {\bibinfo {volume} {66}},\ \bibinfo {pages} {1261} (\bibinfo {year}
  {1997})}\BibitemShut {NoStop}%
\bibitem [{\citenamefont {Shimada}, \citenamefont {Ono},\ and\ \citenamefont
  {Ootuka}(1998)}]{shimada}%
  \BibitemOpen
  \bibfield  {author} {\bibinfo {author} {\bibfnamefont {H.}~\bibnamefont
  {Shimada}}, \bibinfo {author} {\bibfnamefont {K.}~\bibnamefont {Ono}}, \ and\
  \bibinfo {author} {\bibfnamefont {Y.}~\bibnamefont {Ootuka}},\ }\href@noop {}
  {\bibfield  {journal} {\bibinfo  {journal} {J. Phys. Soc. Jpn}\ }\textbf
  {\bibinfo {volume} {67}},\ \bibinfo {pages} {1359} (\bibinfo {year}
  {1998})}\BibitemShut {NoStop}%
\bibitem [{\citenamefont {Ono}, \citenamefont {Shimada},\ and\ \citenamefont
  {Ootuka}(1998)}]{ono2}%
  \BibitemOpen
  \bibfield  {author} {\bibinfo {author} {\bibfnamefont {K.}~\bibnamefont
  {Ono}}, \bibinfo {author} {\bibfnamefont {H.}~\bibnamefont {Shimada}}, \ and\
  \bibinfo {author} {\bibfnamefont {Y.}~\bibnamefont {Ootuka}},\ }\href@noop {}
  {\bibfield  {journal} {\bibinfo  {journal} {{J. Phys. Soc. Jpn}}\ }\textbf
  {\bibinfo {volume} {{67}}},\ \bibinfo {pages} {{2852}} (\bibinfo {year}
  {{1998}})}\BibitemShut {NoStop}%
\bibitem [{\citenamefont {van~der Molen}, \citenamefont {Tombros},\ and\
  \citenamefont {van Wees}(2006)}]{molen}%
  \BibitemOpen
  \bibfield  {author} {\bibinfo {author} {\bibfnamefont {S.~J.}\ \bibnamefont
  {van~der Molen}}, \bibinfo {author} {\bibfnamefont {N.}~\bibnamefont
  {Tombros}}, \ and\ \bibinfo {author} {\bibfnamefont {B.~J.}\ \bibnamefont
  {van Wees}},\ }\href@noop {} {\bibfield  {journal} {\bibinfo  {journal}
  {Phys. Rev. B}\ }\textbf {\bibinfo {volume} {73}},\ \bibinfo {pages} {220406}
  (\bibinfo {year} {2006})}\BibitemShut {NoStop}%
\bibitem [{\citenamefont {Bernand-Mantel}\ \emph {et~al.}(2009)\citenamefont
  {Bernand-Mantel}, \citenamefont {Seneor}, \citenamefont {Bouzehouane},
  \citenamefont {Fusil}, \citenamefont {Deranlot}, \citenamefont {Petroff},\
  and\ \citenamefont {Fert}}]{mantel}%
  \BibitemOpen
  \bibfield  {author} {\bibinfo {author} {\bibfnamefont {A.}~\bibnamefont
  {Bernand-Mantel}}, \bibinfo {author} {\bibfnamefont {P.}~\bibnamefont
  {Seneor}}, \bibinfo {author} {\bibfnamefont {K.}~\bibnamefont {Bouzehouane}},
  \bibinfo {author} {\bibfnamefont {S.}~\bibnamefont {Fusil}}, \bibinfo
  {author} {\bibfnamefont {C.}~\bibnamefont {Deranlot}}, \bibinfo {author}
  {\bibfnamefont {F.}~\bibnamefont {Petroff}}, \ and\ \bibinfo {author}
  {\bibfnamefont {A.}~\bibnamefont {Fert}},\ }\href@noop {} {\bibfield
  {journal} {\bibinfo  {journal} {{Nature Phys.}}\ }\textbf {\bibinfo {volume}
  {{5}}},\ \bibinfo {pages} {{920}} (\bibinfo {year} {{2009}})}\BibitemShut
  {NoStop}%
\bibitem [{\citenamefont {N\'eel}(1949)}]{neel}%
  \BibitemOpen
  \bibfield  {author} {\bibinfo {author} {\bibfnamefont {L.}~\bibnamefont
  {N\'eel}},\ }\href@noop {} {\bibfield  {journal} {\bibinfo  {journal} {Ann.
  G\'eophys}\ }\textbf {\bibinfo {volume} {5}},\ \bibinfo {pages} {99}
  (\bibinfo {year} {1949})}\BibitemShut {NoStop}%
\bibitem [{\citenamefont {Brown}(1963)}]{brown}%
  \BibitemOpen
  \bibfield  {author} {\bibinfo {author} {\bibfnamefont {W.~F.}\ \bibnamefont
  {Brown}},\ }\href@noop {} {\bibfield  {journal} {\bibinfo  {journal} {Phys.
  Rev.}\ }\textbf {\bibinfo {volume} {130}},\ \bibinfo {pages} {1667} (\bibinfo
  {year} {1963})}\BibitemShut {NoStop}%
\end{thebibliography}

\section*{Supplementary Material}
\begin{figure*}[hbp]
\includegraphics[width=0.98\textwidth]{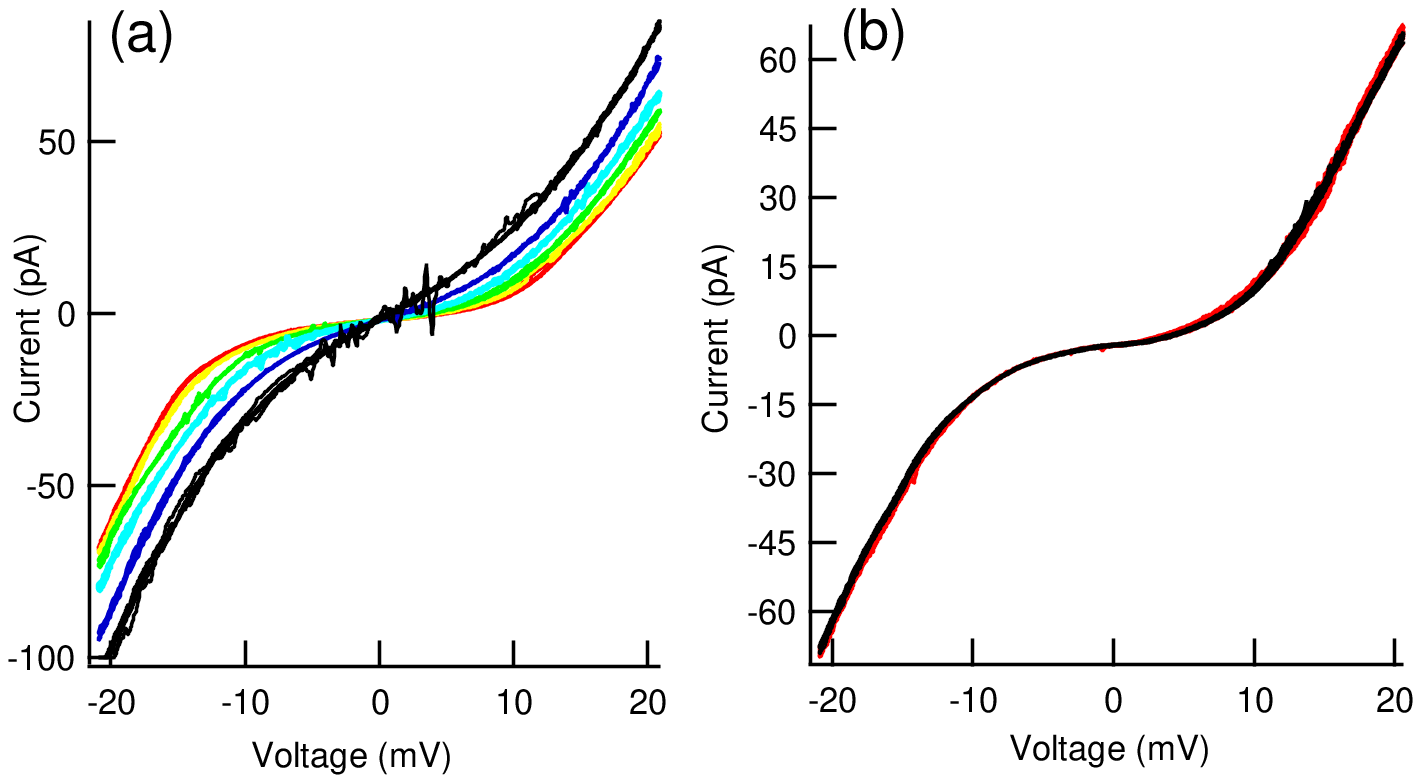}
%\caption{Raw data for Fig. 2(a) Average I-V curves at  $27$K, $16.6$K, $13.1$K, $9.3$K, $5.8$K, and $4.2$K (from top to bottom at positive bias voltage).  (b) Average I-V curves at $T_{s}=$ $1000\mu$s (red) and $1.25\mu$s (black).}
\end{figure*}
Raw data for Fig. 2. (a) All I-V curves at  $27$K, $16.6$K, $13.1$K, $9.3$K, $5.8$K, and $4.2$K (from top to bottom at positive bias voltage).  (b) All I-V curves at $T_{s}=$ $1000\mu$s (red) and $1.25\mu$s (black).

\newpage
\begin{figure}
\includegraphics[width=0.98\textwidth]{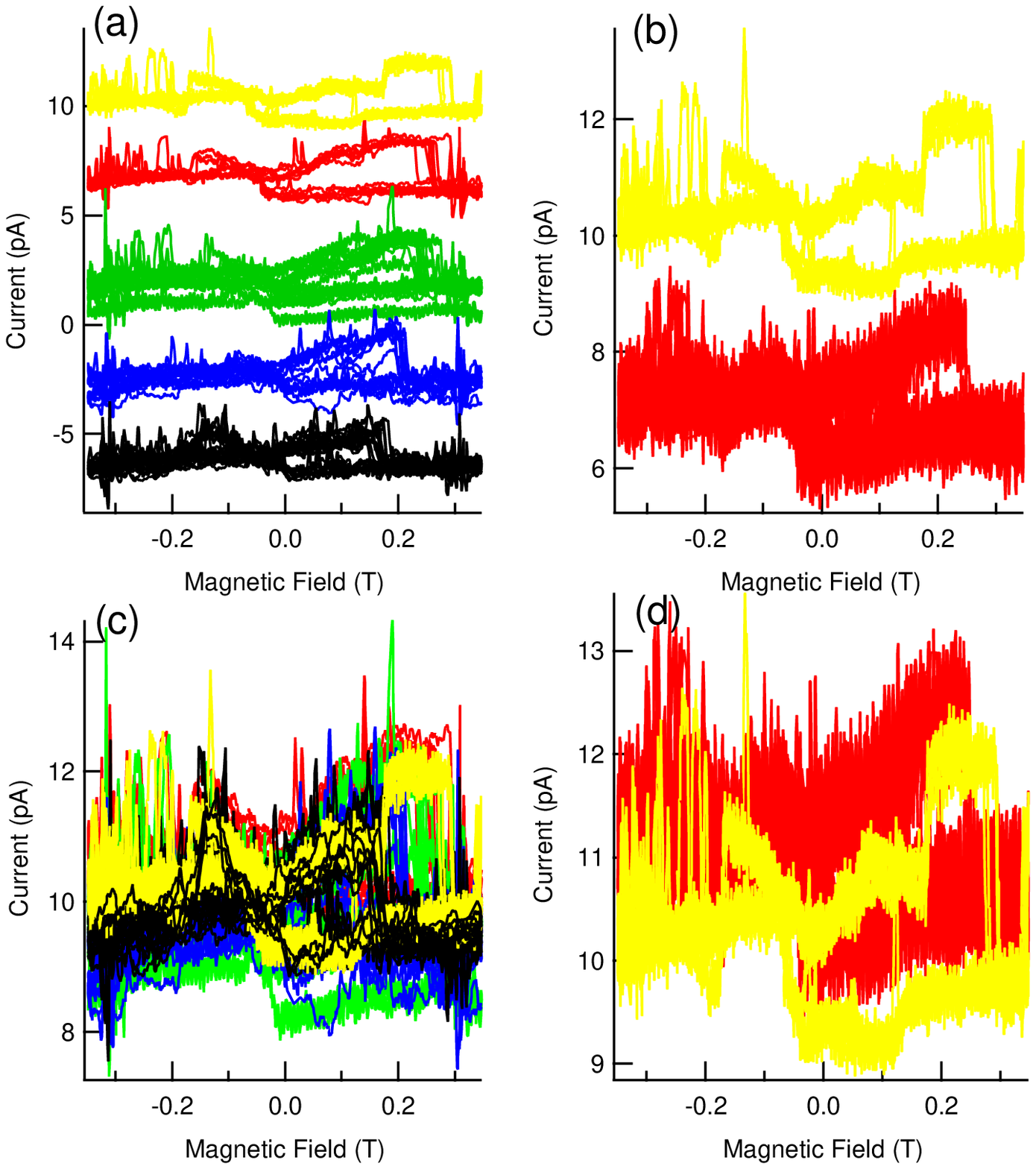}
\end{figure}
Raw data for Fig. 3. (a) All current loops at $4.2$K when $T_{s}=\infty$ (no microwave), $10\mu$s, $5\mu$s, $2.5\mu$s and $1.25\mu$s (top to bottom). (b) All current loops at $4.2$K and $7$K(top to bottom). The loops with finite $T_{s}$ in (a) and at $7$K in (b) are offset with $4$pA spacing. (c) Data of (a) before offset. (d) Data of (b) before offset.
\end{document}